\documentclass[aps,prc,superscriptaddress,twoside,twocolumn,nofootinbib,%
showpacs,floatfix,dvips]{revtex4}
\usepackage{amsmath,amssymb,graphicx,bm}

\begin{document}



\title{Nucleon and $\bm\Delta$ resonances in
$\bm{K \Sigma(1385)}$ photoproduction from nucleons}


\author{Yongseok Oh}%
\email{yoh@comp.tamu.edu}

\affiliation{Cyclotron Institute and Physics Department,
Texas A\&M University, College Station, Texas 77843, U.S.A.}

\author{Che Ming Ko}%
\email{ko@comp.tamu.edu}

\affiliation{Cyclotron Institute and Physics Department,
Texas A\&M University, College Station, Texas 77843, U.S.A.}

\author{K. Nakayama}%
\email{nakayama@uga.edu}

\affiliation{Department of Physics and Astronomy, University of Georgia,
Athens, GA 30602, U.S.A.}
\affiliation{Institut f\"ur Kernphysik, Forschungszentrum
J\"ulich, D-52425 J\"ulich, Germany}



\begin{abstract}

The reaction mechanisms for $K\Sigma(1385)$ photoproduction from the
reaction $\gamma p \to K^+\Sigma^{0}(1385)$ in the resonance energy
region are investigated in a hadronic model. Both contributions from
$N$ and $\Delta$ resonances of masses around $2$~GeV as given in the
Review of Particle Data Group and by the quark model predictions are
included. The Lagrangians for describing the decays of these
resonances into $K\Sigma(1385)$ are constructed with the coupling
constants determined from the decay amplitudes predicted by a quark
model. Comparing the resulting total cross section for the reaction
$\gamma p \to K^+\Sigma^{0}(1385)$ with the preliminary data from
the Thomas Jefferson National Accelerator Facility, we find that the
most important contributions are from the two-star rated resonances
$\Delta(2000) F_{35}$, $\Delta(1940) D_{33}$, and $N(2080) D_{13}$,
as well as the missing resonance $N\frac32^-(2095)$ predicted in the
quark model. Predictions on the differential cross section and
photon asymmetry in this reaction are also given.

\end{abstract}

\pacs{25.20.Lj, 
      13.60.Le, 
      13.60.Rj, 
      14.20.Gk  
     }

\maketitle

\section{Introduction}

Strangeness production from photon-nucleon reactions has been
extensively studied in recent experiments at electron/photon
accelerator facilities~\cite{CLAS05c,LEPS03b,SAPHIR98,Tohoku07}.
Among the motivations for such studies are to obtain a deeper
understanding of the baryon resonances and to search for the
so-called missing resonances, whose existence is predicted by quark
models but has not been experimentally confirmed. Most of the data
from these experiments are for reactions of kaon photoproduction
which are accompanied by the ground state of $\Lambda$ or $\Sigma$
hyperon, i.e., $\Lambda(1116)$ or $\Sigma(1193)$. Recently, there
have been reports on experimental studies of other strangeness
production processes that include $K^* \Lambda$, $K^*\Sigma$, and
$K\Sigma(1385)$ photoproduction~\cite{CLAS06e,CLAS07a,CBELSA}.
Although the reported cross sections for these reactions are smaller
than those for $K\Lambda(1116)$ and $K\Sigma(1193)$ photoproduction,
the suppression factor is not large. In fact, the magnitude of the
cross sections for these reactions in the resonance region,
corresponding to total center-of-mass energies around $2$~GeV, is as
large as one half of the $K\Lambda(1116)$ and $K\Sigma(1193)$
photoproduction cross sections. This indicates that these reaction
channels cannot be neglected in a full coupled channel calculation
for extracting the properties of these baryon
resonances~\cite{CLAS06e}. In addition, these reactions have their
own interesting physics regarding the structure of hadrons. For
example, photoproduction of $K^*\Lambda$ and $K^*\Sigma$ can be used
to obtain information on the properties of strange scalar $\kappa$
meson~\cite{OK06a,OK06b}.

Regarding the missing resonance problem, photoproduction of
$K\Sigma(1385)$ provides a useful tool for testing baryon models
in the literature. According to the quark model of
Ref.~\cite{CR98b}, most nucleon and $\Delta$ resonances have small
couplings to the $K\Sigma(1385)$ channel. Some resonances, mostly
missing or not-well-established ones, are, however, predicted to
have large partial decay widths into this channel. For example, the
missing resonance $N\frac32^-(2095)$ was predicted to have a decay
width of $\Gamma\bm{(} N\frac32^-(2095) \to K\Sigma(1385) \bm{)}
\approx 60$~MeV. Therefore, photoproduction of $K\Sigma(1385)$ could
be an ideal reaction in which one can search for such resonances.

Experimental studies of $K\Sigma(1385)$ photoproduction are very rare,
and only limited experimental data on the total cross section for
$\gamma p \to K^+ \Sigma^0(1385)$ with large error bars have been
reported~\cite{CBCG67,DBCG67,ABBH69}. The CLAS Collaboration at the
Thomas Jefferson National Accelerator Facility recently measured the
cross section of this reaction at 23 different photon energies
covering from the threshold up to $3.8$~GeV~\cite{CLAS06e}. More
accurate data for the total and differential cross sections are
expected to be reported soon~\cite{Guo}. We will use the preliminary
data for the total cross section of this reaction reported in
Ref.~\cite{CLAS06e} for our study.

Theoretical investigation of $K\Sigma(1385)$ photoproduction is also
very scarce. To our knowledge, only a few theoretical studies on
this reaction were reported quite recently. In Ref.~\cite{LS05},
contributions from the single and double $K$-meson pole terms
to the differential cross section of this reaction were compared,
while the role of $\Delta(1700)$ resonance near the threshold
region was addressed in Ref.~\cite{DOS05}. In this paper, we
present a model for $K\Sigma(1385)$ photoproduction from the
reaction $\gamma p \to K^+ \Sigma^0(1385)$, based on an effective
Lagrangian approach. In addition to the $t$-channel $K$ and $K^*$
meson exchanges, we consider the $s$- and $u$-channel diagrams as
well as the contact term, which are required by crossing symmetry
and the gauge invariance condition. We also investigate the role of
resonances in this reaction. For this purpose
we construct the Lagrangians involving the decay
of resonances into $K\Sigma(1385)$. The coupling constants in these
Lagrangians are determined by decomposing the decay amplitudes
according to the relative orbital angular momentum of the final
$K\Sigma(1385)$ state and then comparing them with those known
empirically or calculated from hadronic models.

This paper is organized as follows. In Sec.~\ref{sec:model}, we
discuss the effective Lagrangians employed in the present work. This
includes the general form for the $K^* N \Sigma^*$ interactions (or
$\rho N \Delta$ interactions), which has been overlooked in the
literature. Numerical results on the total and differential cross
sections as well as the photon asymmetry are presented and discussed
in Sec.~\ref{sec:results}, which is followed by a summary and
discussion in Sec.~\ref{sec:conclusion}. The propagators of spin-3/2
and -5/2 baryons are given in Appendix~\ref{app:prop} together with
the isospin structure of the interaction Lagrangians. Given in
Appendix~\ref{app:amp} are details on the decay amplitudes of baryon
resonances into $K\Sigma(1385)$ and $N\gamma$, which are used to
relate the coupling constants in the interaction Lagrangians to the
predicted decay amplitudes from hadronic models.

\section{The model}\label{sec:model}

\subsection{Effective Lagrangians}

\begin{figure}[tb]
\centering
\includegraphics[width=0.8\hsize,angle=0]{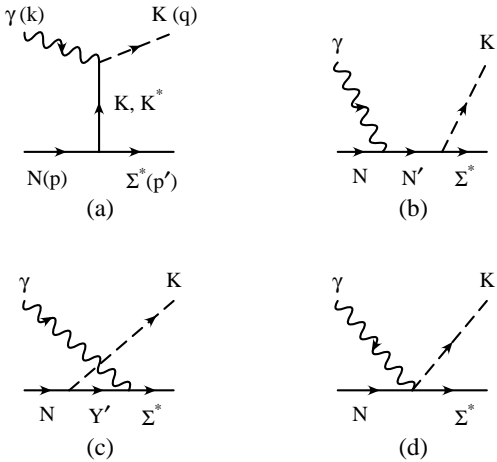}
\caption{Feynman diagrams for $\gamma p \to K^+ \Sigma^{*0}$. $N'$
stands for the nucleon, $\Delta$, and their resonances, and $Y'$ the
$\Lambda$, $\Sigma$, and their resonances. For details of the
contact term [diagram (d)] used in the present work, see
Eq.~(\ref{eq:contact}). } \label{fig:diagram}
\end{figure}

Particle production in photon-nucleon interactions has been
extensively studied in hadronic models based on effective
Lagrangians. This includes the production of various
mesons~\cite{OL04}, charmed hadrons~\cite{LLK03}, $\Xi$
baryons~\cite{NOH06}, and exotic baryons~\cite{exotic1,LK03}. In
this paper, we use this approach to study the reaction $\gamma p \to
K^+ \Sigma^0(1385)$. The production mechanisms for this reaction are
shown in Fig.~\ref{fig:diagram}. Figure~\ref{fig:diagram}(a)
includes the $t$-channel $K$ and $K^*$ meson exchange diagrams. The
$s$-channel diagrams shown in Fig.~\ref{fig:diagram}(b) contain
contributions from non-strange baryons, i.e., nucleon, $\Delta$ and
their resonances. In the present work, we consider resonances of
masses around $2$~GeV as the purpose of this work is to investigate
the role of such resonances in this reaction. Resonances below the
$K\Sigma(1385)$ threshold are not considered since there is no
information on their couplings to the $K\Sigma(1385)$ channel. The
$u$-channel diagrams shown in Fig.~\ref{fig:diagram}(c) contain
hyperons and their resonances. Although all $\Lambda$ and $\Sigma$
resonances can contribute to the reaction through this diagram, only
$\Lambda(1116)$ and $\Sigma(1385)$ are considered in the present
study as there is no information on the photo-transitions between
$\Sigma(1385)$ and hyperon resonances with masses around 2~GeV.
Figure~\ref{fig:diagram}(d) is the contact diagram required by gauge
invariance.

The production amplitudes from the diagrams for $t$-channel $K$
exchange, $s$-channel nucleon term, $u$-channel $\Sigma(1385)$ term,
and the contact term can be calculated from following effective
Lagrangians:
\begin{eqnarray}
\mathcal{L}_{\gamma KK} &=& ie A_\mu \left( K^- \partial^\mu K^+ -
\partial^\mu K^- K^+ \right),
\nonumber \\
\mathcal{L}_{KN\Sigma^*} &=& \frac{f_{KN\Sigma^*}}{M_K} \partial_\mu
\overline{K} \overline{\bm \Sigma}^{*\mu} \cdot \bm{\tau} N + \mbox{
H.c.},
\nonumber \\
\mathcal{L}_{\gamma NN} &=& -e \bar{N} \left( \gamma^\mu A_\mu
\frac{1+\tau_3}{2} - \frac{\kappa_N^{}}{2M_N} \sigma^{\mu\nu}
\partial_\nu A_\mu \right) N,
\nonumber \\
\mathcal{L}_{\gamma KN\Sigma^*} &=& -ie \frac{f_{KN\Sigma^*}}{M_K}
A^\mu K^- \left( \bar{\Sigma}_\mu^{*0} p + \sqrt{2}
\bar{\Sigma}_\mu^{*+} n \right)
\nonumber \\ && \mbox{}
+ \mbox{ H.c.},
\nonumber \\
\mathcal{L}_{\gamma\Sigma^*\Sigma^*} &=& e \overline{\Sigma}_\mu^*
A_\alpha \Gamma_{\gamma\Sigma^*}^{\alpha,\mu\nu} \Sigma_\nu^*,
\label{eq:lag1}
\end{eqnarray}
where $M_K$ is the kaon mass, $A_\mu$ is the photon field, and
$\Sigma^*_\mu$ is the Rarita-Schwinger field for the $\Sigma(1385)$
of spin-$3/2$.%
\footnote{In this work, we do not consider the off-shell properties of
the Rarita-Schwinger field.}
The isodoublets are defined as
\begin{equation}
K = \left( \begin{array}{c} K^+ \\ K^0  \end{array} \right), \qquad
\overline{K} = \left( K^-, \bar{K}^0 \right), \qquad
N = \left( \begin{array}{c} p \\ n  \end{array} \right).
\label{eq:isodoublet}
\end{equation}
The electromagnetic interaction of the $\Sigma^*$ field contains
\begin{eqnarray}
A_\alpha \Gamma_{\gamma\Sigma^*}^{\alpha,\mu\nu} &=& \left\{
g^{\mu\nu} \gamma^\alpha - \frac12 \left( \gamma^\mu \gamma^\nu
\gamma^\alpha + \gamma^\alpha \gamma^\mu \gamma^\nu \right) \right\}
A_\alpha T_3
\nonumber \\ && \mbox{}
- \frac{\kappa_{\Sigma^*}^{}}{2M_N} \sigma^{\alpha\beta} \partial_\beta
A_\alpha g^{\mu\nu},
\end{eqnarray}
where $M_N$ is the nucleon mass and $T_3 = \mbox{diag}(1,0,-1)$.

For the $KN\Sigma^*$ coupling, it can be related to the $\pi N
\Delta$ coupling by the SU(3) flavor symmetry relation,
\begin{equation}
\frac{f_{\pi N \Delta}}{M_\pi} = -\sqrt6 \frac{f_{KN\Sigma^*}}{M_K},
\end{equation}
where $M_\pi$ is the pion mass. Estimating the $\pi N \Delta$
coupling as $f_{\pi N \Delta} = 2.23$ from the Delta resonance decay
width $\Gamma(\Delta \to N\pi) = 120$ MeV, we obtain from the above
equation $f_{KN\Sigma^*} = -3.22$. The electromagnetic interactions
of baryons contain the baryon anomalous magnetic moments. We use the
empirical value $\kappa_p^{} = 1.793$ for the proton. Since the
magnetic moment of $\Sigma^0(1385)$ is unknown, its value is taken
from the quark model prediction given in Ref.~\cite{Lic77}, i.e.,
$\kappa_{\Sigma^{*0}}^{} = 0.36$.

For the $t$-channel $K^*$ exchange, we use the Lagrangian,
\begin{equation}
\mathcal{L}_{\gamma KK^*} =
g^{}_{K^*K\gamma} \varepsilon^{\mu\nu\alpha\beta} \partial_\mu A_\nu
\partial_\alpha K_\beta^{*-} \bar{K}^+ + \mbox{ H.c.},
\end{equation}
for the $\gamma KK^*$ interaction with $g^{}_{K^*K\gamma} =
0.254$~GeV$^{-1}$, which is determined from the empirical value of
the $K^*$ decay width $\Gamma(K^{*\pm} \to K^\pm \gamma) \approx
50$~keV. We note that for neutral $K^*$, $g^{}_{K^*K\gamma} =
-0.388$~GeV$^{-1}$ as $\Gamma(K^{*0} \to K^0 \gamma) \approx
116$~keV, and the signs of these coupling constants are fixed by the
quark model.

For the interactions of a vector meson with spin-1/2 and spin-3/2
baryons, i.e., for the vertex of $\frac32 \to 1 + \frac12$, there
are in general three independent interaction terms from
consideration of angular momentum and parity conservation. The most
general form of the $K^*N\Sigma^*$ interaction Lagrangian can be
written as
\begin{eqnarray}
\mathcal{L}_{K^*N\Sigma^*} &=& \frac{ig_1^{}}{2M_N}
\overline{K}^{*\mu\nu} \overline{\bm\Sigma}^*_\mu \cdot \bm{\tau}
\gamma_\nu \gamma_5^{} N
\nonumber \\ && \mbox{}
+ \frac{g_2^{}}{(2M_N)^2} \overline{K}^{*\mu\nu}
\overline{\bm\Sigma}^*_\mu \cdot \bm{\tau} \gamma_5^{} \partial_\nu N
\nonumber \\ && \mbox{}
- \frac{g_3^{}}{(2M_N)^2}
\partial_\nu \overline{K}^{*\mu\nu} \overline{\bm\Sigma}^*_\mu \cdot
\bm{\tau} \gamma_5^{} N + \mbox{ H.c.},
\end{eqnarray}
where $K^*_{\mu\nu} = \partial_\mu K^*_\nu - \partial_\nu K^*_\mu$
and $K^*$ is an isodoublet as $K$ in Eq.~(\ref{eq:isodoublet}). To
determine the coupling constants $g_{1,2,3}^{}$, we again make use
of the SU(3) relations to relate them to the $\rho N \Delta$
coupling. For the coupling constant $g_1$, the SU(3) relation
\begin{equation}
\frac{g_1^{\rho N \Delta}}{M_\rho} = -\sqrt6 \frac{g_1^{K^* N
\Sigma^*}}{2M_N}
\end{equation}
leads to $g_1^{} = -5.48$ for the $K^* N \Sigma^*$ coupling if the
empirically determined value $g_1^{\rho N \Delta} =
5.5$~\cite{KA04,ONL04} is used. Since the other two couplings,
$g_2^{}$ and $g_3^{}$, in the $\rho N \Delta$ interactions have
never been seriously considered in previous studies, corresponding
couplings for the $K^* N \Sigma^*$ interactions thus cannot be
determined. In the present study, we treat $g_2^{}$ and $g_3^{}$ in
the $K^* N \Sigma^*$ interactions as free parameters and vary their
values to find their role in $K\Sigma(1385)$ photoproduction.

The $u$-channel diagrams shown in Fig.~\ref{fig:diagram}(c) contain
intermediate hyperon $Y'$. Because of the lack of information on the
radiative decays of hyperon resonances to $\Sigma(1385)$, we
consider only the contribution from the ground state hyperons. The
effective Lagrangians for these diagrams are
\begin{eqnarray}
\mathcal{L}_{\Sigma^*Y\gamma} &=& -\frac{ief_1}{2M_Y} \overline{Y} \gamma_\nu
\gamma_5 F^{\mu\nu} \Sigma_\mu^*
\nonumber \\ && \mbox{}
- \frac{ef_2}{(2M_Y)^2} \partial_\nu \overline{Y} \gamma_5 F^{\mu\nu}
\Sigma_\mu^* + \mbox{ H.c.},
\nonumber \\
\mathcal{L}_{KNY} &=& \frac{g_{KNY}^{}}{M_N +  M_Y} \overline{N} \gamma^\mu
\gamma_5^{} Y \partial_\mu K + \mbox{ H.c.},
\end{eqnarray}
where $F_{\mu\nu} = \partial_\mu A_\nu - \partial_\nu A_\mu$ and $Y$
stands for a hyperon with spin-$1/2$.
For the intermediate $\Lambda(1116)$ state, we use the radiative decay
width, $\Gamma\bm{(}\Sigma(1385) \to \Lambda \gamma \bm{)} = 479 \pm
120$~keV, as recently measured by CLAS Collaboration.
To estimate the relative strength of the two coupling
constants, we make use of the quark model result of
Ref.~\cite{WPR91} for the helicity amplitudes $A_{1/2}$ and $A_{3/2}$
(see Appendix~\ref{app:amp}),
which gives the ratio $A_{3/2}/A_{1/2} \approx 1.82$.%
\footnote{We note that most theoretical predictions~\cite{WPR91,WBF98}
underestimate this decay width.}
Combining the two information, we obtain
\begin{equation}
f_1 = 4.52, \qquad f_2 = 5.63.
\end{equation}
The coupling constant $g_{KN\Lambda}^{}$  can be determined by
flavor SU(3) symmetry relations, which give $g_{KN\Lambda}^{} =
-13.24$. For the intermediate $\Sigma(1193)$ hyperon, there is no
experimental data for $\Gamma\bm{(}\Sigma(1385) \to \Sigma(1193)
\gamma \bm{)}$. However, most hadron model calculations show that
the decay width of $\Sigma(1385) \to \Sigma(1193) \gamma$ is less
than 10\% of that of $\Sigma(1385) \to \Lambda(1116)
\gamma$~\cite{WPR91,WBF98}.
With $g_{KN\Sigma} = 3.58$ from the flavor SU(3) symmetry, we have
estimated that the contribution from the $u$-channel $\Sigma(1193)$ is
only at the level of 0.5\% of the $u$-channel $\Lambda(1116)$ contribution.
In the present work, therefore, we will consider the $u$-channel
diagram with the $\Lambda(1116)$ hyperon only.

As we have mentioned above, the motivation for the study of
$K\Sigma(1385)$ photoproduction is to identify the contributions
from baryon resonances. For this purpose, we consider the
contributions from both nucleon and $\Delta$ resonances in the
present work, which requires the interaction Lagrangians for
photoexcitation of a resonance from a nucleon as well as for its
decay into $K\Sigma(1385)$ channel. For the former, we use
\begin{eqnarray}
\mathcal{L}_{RN\gamma}(\textstyle\frac12^\pm) &=& \frac{ef_1}{2M_N}
\bar{N} \Gamma^{(\mp)} \sigma_{\mu\nu} \partial^\nu A^\mu R +
\mbox{H.c.},
\nonumber \\
\mathcal{L}_{RN\gamma}(\textstyle\frac32^\pm) &=& -
\frac{ief_1}{2M_N} \overline{N} \Gamma^{(\pm)}_\nu F^{\mu\nu} R_\mu
\nonumber \\ && \mbox{}
- \frac{ef_2}{(2M_N)^2} \partial_\nu \bar{N}
\Gamma^{(\pm)} F^{\mu\nu} R_\mu + \mbox{H.c.},
\nonumber \\
\mathcal{L}_{RN\gamma}(\textstyle\frac52^\pm) &=&
\frac{ef_1}{(2M_N)^2} \bar{N} \Gamma_\nu^{(\mp)}
\partial^\alpha F^{\mu\nu} R_{\mu\alpha}
\nonumber \\ && \mbox{}
-\frac{ief_2}{(2M_N)^3} \partial_\nu \bar{N} \Gamma^{(\mp)}
\partial^\alpha F^{\mu\nu} R_{\mu\alpha}
+ \mbox{H.c.},
\nonumber \\
\label{eq:RNgamma}
\end{eqnarray}
where $R$, $R_\mu$, and $R_{\mu\nu}$ are the fields for the
spin-$1/2$, $3/2$, and $5/2$ resonances, respectively, with
\begin{equation}
\Gamma^{(\pm)}_\mu = \left( \begin{array}{c} \gamma_\mu \gamma_5^{} \\
\gamma_\mu \end{array} \right), \qquad \Gamma^{(\pm)} = \left(
\begin{array}{c} \gamma_5^{} \\
\textbf{1} \end{array} \right).
\label{gamma_pm}
\end{equation}
It should be noted that the coupling constant $f_i$ has isospin
dependence if the resonance $R$ has isospin $1/2$, while it is
isospin blind if the isospin of $R$ is $3/2$. Since we only consider
in the present work photoexcitation of resonances from the proton,
the isospin quantum number is fixed in the process.

For the decay of a resonance with spin $j$ into $K\Sigma(1385)$, the
number of possible interaction terms in the Lagrangian is restricted
by the angular momentum and parity conservation. The interaction
Lagrangian has one term for a $j=\frac12$ resonance but has two
terms for a resonance with $j \ge \frac32$. Explicitly, the
effective Lagrangians for $RK\Sigma^*$ interactions can be written
as
\begin{eqnarray}
\mathcal{L}_{RK\Sigma^*}(\textstyle\frac12^\pm) &=& \frac{h_1}{M_K}
\partial_\mu K \bar{\Sigma}^{*\mu} \Gamma^{(\mp)} R + \mbox{H.c.},
\nonumber \\
\mathcal{L}_{RK\Sigma^*}(\textstyle\frac32^\pm) &=& \frac{h_1}{M_K}
\partial^\alpha K \bar{\Sigma}^{*\mu} \Gamma_\alpha^{(\pm)} R_\mu
\nonumber \\ && \mbox{} + \frac{ih_2}{M_K^2} \partial^\mu
\partial^\alpha K \bar{\Sigma}^*_\alpha \Gamma^{(\pm)} R_\mu +
\mbox{H.c.},
\nonumber \\
\mathcal{L}_{RK\Sigma^*}(\textstyle\frac52^\pm) &=& \frac{i h_1}{M_K^2}
\partial^\mu \partial^\beta K \bar{\Sigma}^{*\alpha} \Gamma_\mu^{(\mp)}
R_{\alpha\beta}
\nonumber \\ && \mbox{}
- \frac{h_2}{M_K^3}\partial^\mu
\partial^\alpha \partial^\beta K \bar{\Sigma}^*_\mu \Gamma^{(\mp)}
R_{\alpha\beta} + \mbox{H.c.}.
\nonumber \\
\label{eq:RKSigma*}
\end{eqnarray}

In evaluating the Feynman diagrams (b) and (c) in
Fig.~\ref{fig:diagram} for the reaction $\gamma p \to
K^+\Sigma^{0}(1385)$, we need also the propagators of baryon
resonances. They are given explicitly in Appendix~\ref{app:prop} for
baryon resonances of spins up to $5/2$, together with the isospin
structure of their interaction Lagrangians. For the coupling
constants $f_{1,2}$ in Eq.~(\ref{eq:RNgamma}) and $h_{1,2}$ in
Eq.~(\ref{eq:RKSigma*}), they can be related to the photon helicity
amplitudes of $R \to N\gamma$ and the decay amplitudes of $R \to
K\Sigma(1385)$, respectively. These relations are given explicitly
in Appendix~\ref{app:amp}, and they allow us to determine the
coupling constants once these amplitudes are known either
empirically or from models for hadrons.

\subsection{Form factors}

In evaluating the production amplitudes of $\gamma p\to K^+
\Sigma^0(1385)$, we need to dress the interaction vertices with form
factors. We use the monopole type form factor for the $t$-channel
$K$ meson exchange diagram, i.e.,
\begin{equation}
F_M(q_{\rm ex}^2,M_{\rm ex}) = \frac{\Lambda_M^2 - M_{\rm
ex}^2}{\Lambda_M^2 - q_{\rm ex}^2}.
\label{eq:ffa}
\end{equation}

For $s$- and $u$-channel diagrams and $t$-channel $K^*$ exchange, we
adopt the form factor
\begin{equation}
F_B(q_{\rm ex}^2,M_{\rm ex}) =
\left(\frac{n\Lambda_B^4}{n\Lambda_B^4 + (q_{\rm ex}^2 - M_{\rm
ex}^2)^2} \right)^n, \label{eq:ffb}
\end{equation}
which goes to a Gaussian form as $n \to \infty$. In
Eqs.~(\ref{eq:ffa}) and (\ref{eq:ffb}), $q_{\rm ex}$ is the
four-momentum of the exchanged particle of mass $M_{\rm ex}$. The
cutoff parameters $\Lambda_M$ and $\Lambda_B$ as well as $n$ will be
adjusted to fit the experimental data.

\subsection{Generalized contact current}

Employing different form factors to interaction vertices breaks
gauge invariance. Following the prescription of Ref.~\cite{HNK06},
we restore gauge invariance by introducing the following generalized
contact term to the amplitude for the reaction $\gamma p\to
K^+\Sigma^0(1385)$:
\begin{equation}
M_c^{\mu\nu}= \Gamma^{\mu\nu}_{\gamma KN\Sigma^*} f_t
+ie\Gamma^\nu_{KN\Sigma^*}(q) C^\mu.
\label{eq:contact}
\end{equation}
In the above equation,
\begin{equation}
\Gamma^{\mu\nu}_{\gamma KN\Sigma^*} = ie \frac{f_{KN\Sigma^*}}{M_K}
g^{\mu\nu}
\end{equation}
and
\begin{equation}
\Gamma^\nu_{KN\Sigma^*}(q) = - \frac{f_{KN\Sigma^*}}{M_K} q^\nu,
\end{equation}
are vertex functions obtained, respectively, from $\mathcal{L}_{\gamma
KN\Sigma^*}$ and $\mathcal{L}_{KN\Sigma^*}$ in Eq.~(\ref{eq:lag1})
with $q$ being the momentum of the outgoing $K$ meson;
$C^\mu$ is defined as
\begin{eqnarray}
C^\mu &=& - (2q-k)^\mu \frac{f_t - 1}{t - M_K^2} f_s - (2p+k)^\mu
\frac{f_s-1}{s-M_N^2} f_t,
\nonumber \\
\label{eq:cmu}
\end{eqnarray}
with the momenta defined in Fig.~\ref{fig:diagram}, and the form
factors in $t$-channel [Eq.~(\ref{eq:ffa})] and $s$-channel
[Eq.~(\ref{eq:ffb})] diagrams are denoted by $f_t$ and $f_s$,
respectively. Note that the first term on the right-hand-side of
Eq.~(\ref{eq:contact}) is the usual Kroll-Ruderman contact current
multiplied by the $t$-channel form factor. The last term is an
additional contact current required to restore gauge invariance of
the total amplitude in the presence of form factors at the hadronic
vertices.
We note that the contact current as specified above 
also satisfies the crossing symmetry.
For details on the restoration of gauge invariance and a
more general form of $C^\mu$, we refer the readers to
Ref.~\cite{HNK06}.

\subsection{$\bm{N}$ and $\bm{\Delta}$ resonances}

\begin{table*}[tbh]
\centering
\begin{tabular}{c|c|cc|cc|cc|cc}
\hline\hline Resonance & PDG~\cite{PDG06} &
\multicolumn{2}{c|}{Amplitudes of $R \to K\Sigma(1385)$ ${}^\dagger$
}&
$h_1$ & $h_2$ & \multicolumn{2}{c|}{Amplitudes of $R \to N\gamma$
${}^{\dagger\dagger}$ }&
$f_1$ & $f_2$ \\
 & & $G(\ell_1)$ & $G(\ell_2)$ & & & $A_{1/2}^p$ & $A_{3/2}^p$ & & \\ \hline
$N\frac12^-(1945)$ & $S_{11}^{*}(2090)$ & $G(2) = +1.7$ & --- &
$-9.8$  & --- & $+12$ & --- & $-0.055$  & ---
\\
$N\frac32^-(1960)$ & $ D_{13}^{**}(2080)$ & $G(0) = +1.3$ & $G(2) =
+1.4$ & $0.24$ & $-0.54$ &  $+36$  & $-43$ & $-1.25$ & $1.21$
\\
$N\frac52^-(2095)$ & $ D_{15}^{**}(2200)$ & $G(2) = -2.0$ & $G(4) =
0.0$ & $0.29$ & $-0.08$ &  $-9$  & $-14$ & $0.37$ & $-0.57$
\\
$\Delta\frac32^-(2080)$ & $ D_{33}^{*}(1940)$ & $G(0) = -4.1$ &
$G(3) = -0.5$ & $-0.68$ & $1.00$ &  $-20$ & $-6$  & $0.39$ & $-0.57$
\\
$\Delta\frac52^+(1990)$ & $ F_{35}^{**}(2000)$ & $G(1) = +4.0$ &
$G(3) = -0.1$ & $-0.87$ & $0.11$ &  $-10$ & $-28$  & $-0.68$ &
$-0.062$
\\
\hline\hline
\end{tabular}

${}^\dagger$: in $\sqrt{\mbox{GeV}}$ \\
${}^{\dagger\dagger}$ in $10^{-3}/\sqrt{\mbox{GeV}}$

\caption{Resonances listed in the review of PDG~\cite{PDG06} and
their decay amplitudes of $R \to K\Sigma(1385)$ and of $R \to
N\gamma$ predicted in Refs.~\cite{CR98b,Caps92}. The coupling
constants are calculated using the resonance masses of PDG.}
\label{tab:res-A}
\end{table*}

For resonances in the $s$-channel diagrams, we include those with
spin $j\le$ $5/2$. Neglecting resonances with higher spins is
justified as they have been shown in Ref.~\cite{CR98b} to couple
weakly to the $K\Sigma(1385)$ channel. We classify the resonances
into two groups: (A) resonances listed in the review of Particle
Data Group (PDG)~\cite{PDG06} and (B) missing resonances. Since the
decay widths of the resonances listed in PDG into $K\Sigma(1385)$
have not been empirically determined, we have to rely on theoretical
models, such as the quark model of Refs.~\cite{Caps92,CR98b}, to
determine their coupling constants to $K\Sigma(1385)$. Instead of
all possible resonances, we consider only a few of them which are
predicted to have large couplings to $N\gamma$ and to
$K\Sigma(1385)$. The resonances of group (A) and the predictions of
the quark model given in Refs.~\cite{CR98b,Caps92} on their decay
amplitudes are given in Table~\ref{tab:res-A}. These resonances are
referred to as PDG resonances and include $N\frac12^-(1945)$,
$N\frac32^-(1960)$, $N\frac52^-(2095)$, $\Delta\frac32^-(2080)$, and
$\Delta\frac52^+(1990)$, which are identified as $N(2090) S_{11}^*$,
$N(2080) D_{13}^{**}$, $N(2200) D_{15}^{**}$, $\Delta(1940)
D_{33}^*$, and $\Delta(2000) F_{35}^{**}$, respectively, by the
authors of Ref.~\cite{CR98b}. Although listed in the review of PDG,
these resonances are rated as either one-star or two-star
resonances, which means that the evidence of their existence is poor
or only fair~\cite{PDG06} and that further works are required to
verify their existence and to know their properties. Accordingly,
their total decay widths and branching ratios are not known. In the
present work, we assume the same total decay width of $\Gamma_R =
300$~MeV for these resonances.

\begin{table*}[tbh]
\centering
\begin{tabular}{c|cc|cc|cc|cc}
\hline\hline Resonance & \multicolumn{2}{c|}{Amplitudes of $R \to
K\Sigma(1385)$ ${}^\dagger$ }& $h_1$ & $h_2$ &
\multicolumn{2}{c|}{Amplitudes of $R \to N\gamma$
${}^{\dagger\dagger}$ }&
$f_1$ & $f_2$ \\
 & $G(\ell_1)$ & $G(\ell_2)$ & & & $A_{1/2}^p$ & $A_{3/2}^p$ & & \\ \hline
$N\frac32^-(2095)$ & $G(0) = +7.7$ & $G(2) = -0.8$ & $0.99$ & $0.27$
& $-9$ & $-14$ & $0.49$ & $-0.83$
\\
$N\frac52^+(1980)$ & $G(1) = -3.6$ & $G(3) = -0.1$ & $0.59$ & $0.24$
& $-11$ & $-6$ & $0.019$ & $-0.13$
\\
$\Delta\frac32^-(2145)$ & $G(0) = +5.2$ & $G(2) = -1.9$ & $0.25$ &
$0.46$ & $0$ & $+10$ & $0.11$ & $-0.059$
\\
\hline\hline
\end{tabular}

${}^\dagger$: in $\sqrt{\mbox{GeV}}$ \\
${}^{\dagger\dagger}$ in $10^{-3}/\sqrt{\mbox{GeV}}$

\caption{Missing resonances and their decay amplitudes predicted in
Refs.~\cite{CR98b,Caps92}.} \label{tab:res-B}
\end{table*}

The missing resonances which are predicted to have large couplings
to the $K\Sigma(1385)$ channel are listed in Table~\ref{tab:res-B}.
These resonances include $N\frac32^-(2095)$, $N\frac52^+(1980)$, and
$\Delta\frac32^-(2145)$. Among them the resonance $N\frac32^-(2095)$
is particularly interesting since it is predicted to have a very
large decay width into the $K\Sigma(1385)$ channel,
$\Gamma\bm{(}N\frac32^-(2095)\to K\Sigma(1385) \bm{)} \simeq
60$~MeV~\cite{CR98b}. Photoproduction of $K\Sigma(1385)$ thus offers
an opportunity for finding these resonances. In Ref.~\cite{CR98b},
the missing resonances $N\frac12^-(2070)$ and $N\frac52^-(2260)$ are
also predicted to have large couplings to $K\Sigma(1385)$, but no
prediction for their photoexcitation amplitudes have been made
within the same model. We thus leave the investigation on the role
of these resonances to a future study. We also assume $\Gamma_R =
300$ MeV for these resonances.

\section{Results} \label{sec:results}

\subsection{Total cross section}

\begin{figure}[tb]
\centering
\includegraphics[width=1.0\hsize,angle=0]{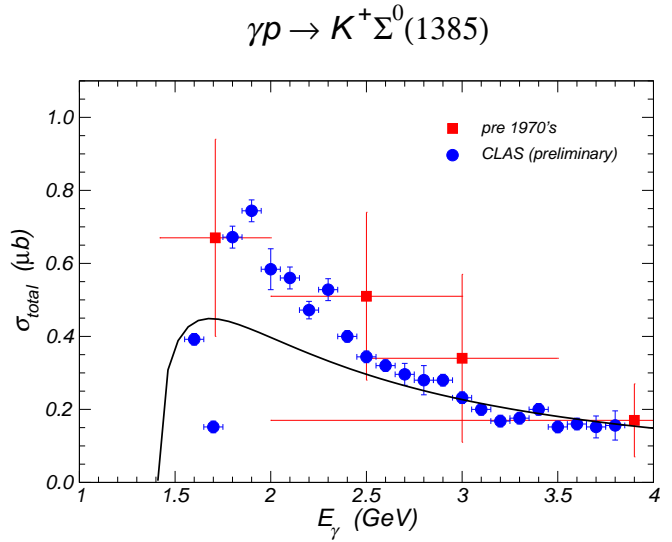}
\caption{(Color online) Total cross sections for $\gamma p \to K^+
\Sigma^0(1385)$ without resonance contributions. The pre-1970s data
are from Refs.~\cite{CBCG67,DBCG67,ABBH69} and the preliminary data
of CLAS Collaboration are from Ref.~\cite{CLAS06e}.}
\label{fig:total-1}
\end{figure}

With the effective Lagrangians and form factors constructed above,
we first compute the total cross section for $\gamma p \to K^+
\Sigma^0(1385)$ without the resonance contributions.
For the form factors in the $s$- and $u$-channel diagrams, we take
$\Lambda_B = 1.0$~GeV with $n=1$. For the $t$-channel $K$ exchange, we use
$\Lambda_M = 0.83$~GeV to reproduce the total cross section data at
$E_\gamma \ge 2.5$~GeV. Following Ref.~\cite{MHE87}, we avoid the
use of $F_M$ for vector meson exchanges and the $t$-channel $K^*$
exchange is calculated by using the form factor $F_B$ with
$\Lambda_B=1.2$~GeV and $n=1$. With the $K^*N\Sigma^*$ coupling
constants determined before, namely, $g_1^{} = -5.48$ and $g_2^{} =
g_3^{} = 0$, we find that the contribution from the $K^*$ exchange
is negligible in the considered energy region. Even at higher
energies, $E_\gamma = 3 \sim 4$ GeV, the $K^*$ exchange contribution
is only at the level of a few percent of those from other production
mechanisms. We have also tested the role of the $K^*$ exchange by
allowing non-vanishing values for $g_2^{}$ and $g_3^{}$. We again
find that the $K^*$ exchange is suppressed compared with other
production mechanisms unless $g_2$ and/or $g_3$ is as large as $\sim
100$. Although there is no constraint at present on the values of
$g_2$ and $g_3$, we regard such a large value unrealistic. This
leads us to conclude that the role of $K^*$ exchange in this
reaction is negligibly small. However, since the $K^*$ trajectory
has a larger intercept than the $K$ trajectory, the role of the
$K^*$ exchange would have a chance to be revealed at very high
energies. It is thus of interest to measure the cross sections at
much higher energies, and this would help constrain the values of
the coupling constants $g_2$ and $g_3$.

Our result on the total cross section is shown in Fig.~\ref{fig:total-1}
and is compared with the pre-1970's data~\cite{CBCG67,DBCG67,ABBH69} and
the preliminary CLAS data reported in Ref.~\cite{CLAS06e}.%
\footnote{The preliminary CLAS data give very small cross sections
for $E_\gamma \le 1.7$~GeV, which deviate significantly from our
prediction. These two data points are now corrected in the new
analyses of the CLAS data which are in progress~\cite{Guo}.}
Comparison with the preliminary CLAS data for the total cross
section of $\gamma p \to K \Sigma^0(1385)$ shows that this model can
explain the general energy dependence of the total cross section but
not the enhanced cross section at $E_\gamma = 1.7 \sim 1.9$ GeV.
Although varying the cutoff parameters of employed form factors can
change the magnitude of the cross section, the peak arising from the
threshold effect cannot reproduce the observed peak in the data.
This implies that resonances play an important role in the
production mechanism.

\begin{figure}[tb]
\centering
\includegraphics[width=1.0\hsize,angle=0]{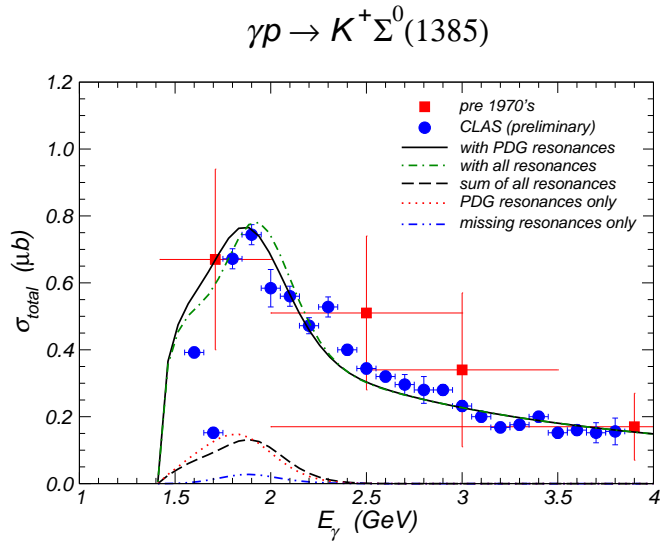}
\caption{(Color online) Total cross sections for the $\gamma p \to
K^+ \Sigma^0(1385)$ reaction with the resonances listed in
Tables~\ref{tab:res-A} and \ref{tab:res-B}. See the text for the
details.} \label{fig:total-res}
\end{figure}

Including the $s$-channel nucleon and $\Delta$ resonances listed in
Tables~\ref{tab:res-A} and \ref{tab:res-B} in the reaction $\gamma p
\to K^+ \Sigma^0(1385)$, we have recalculated its cross section. In
this calculation, the parameters of the non-resonant terms are fixed
as before, while the resonance terms are obtained by using the form
factor $F_B$ in the form of the Gaussian function that is obtained
by taking $n \to \infty$ and the cutoff $\Lambda_B = 1.0$~GeV, as
motivated by the Gaussian radial wave functions in the quark model.
The resulting total cross section for the reaction $\gamma p \to K^+
\Sigma^0(1385)$ is shown in Fig.~\ref{fig:total-res}. As shown by
the dashed line, the contribution from all resonances to the total
cross section of $\gamma p \to K^+ \Sigma^0(1385)$ is important in
the region around $E_\gamma = 1.8 \sim 2.0$~GeV. Although the
contribution coming from the missing resonances (dash-dash-dotted
line) is small compared to the PDG resonance contributions (dotted
line), it moves the peak coming from the resonant terms to a
somewhat higher energy. This result shows that most resonance
contributions come from the sum of the PDG resonances. However, it
should be kept in mind that this conclusion follows from the quark
model predictions of Refs.~\cite{CR98b,Caps92} for the empirically
not well-known decay properties of the PDG resonances. Therefore,
detailed studies on this reaction could be used to constrain the
properties of the PDG resonances listed in Table~\ref{tab:res-A}.
The total cross sections obtained by including only the PDG
resonances and all the resonances considered in the present work,
which also includes the missing resonances predicted by quark
models, are given, respectively, by the solid line and the
dot-dashed line in Fig.~\ref{fig:total-res}. These results show that
the peak observed in the preliminary CLAS data can be successfully
explained by these resonances.

\begin{figure}[t]
\centering
\includegraphics[width=1.0\hsize,angle=0]{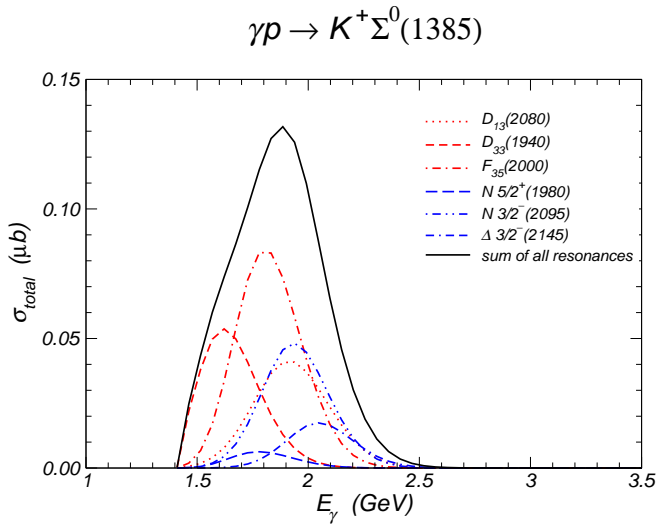}
\caption{(Color online)
Contributions from various resonances to the total cross
section for $\gamma p \to K^+ \Sigma^0(1385)$.}
\label{fig:resonance}
\end{figure}

The contributions from different resonances are shown separately in
Fig.~\ref{fig:resonance}.
The solid line is the sum of all the resonances considered in this
work and, therefore, corresponds to the dashed line in
Fig.~\ref{fig:total-res}.
The largest contribution comes from the $\Delta$ resonance
$\Delta(2000) F_{35}$ (the dot-dashed line in Fig.~\ref{fig:resonance}),
and the contributions from $\Delta(1940) D_{33}$ (the short dashed line
in Fig.~\ref{fig:resonance}) and $N(2080) D_{13}$ (the dotted line in
Fig.~\ref{fig:resonance}) are also noticeable.
One interesting result is that the contribution from the missing
resonance $N\frac32^-(2095)$ is not the dominant one, although it is
as large as that from $N(2080) D_{13}$.
As discussed above, this missing resonance is predicted to have a very
large coupling to $K\Sigma(1385)$.
Its effect in the reaction $\gamma p \to K^+ \Sigma^0(1385)$ is, however,
not large as a result of its rather small couplings to $N\gamma$.
Furthermore, this resonance has a destructive interference with the
other missing resonance, $\Delta\frac32^-(2145)$, so that the net
contribution from missing resonances becomes small.
For $N(2090) S_{11}$ and $N(2200) D_{15}$, their contributions are found
to be too small to be shown in Fig.~\ref{fig:resonance}.

\subsection{Differential cross section and photon asymmetry}

\begin{figure}[tb]
\centering
\includegraphics[width=1.0\hsize,angle=0]{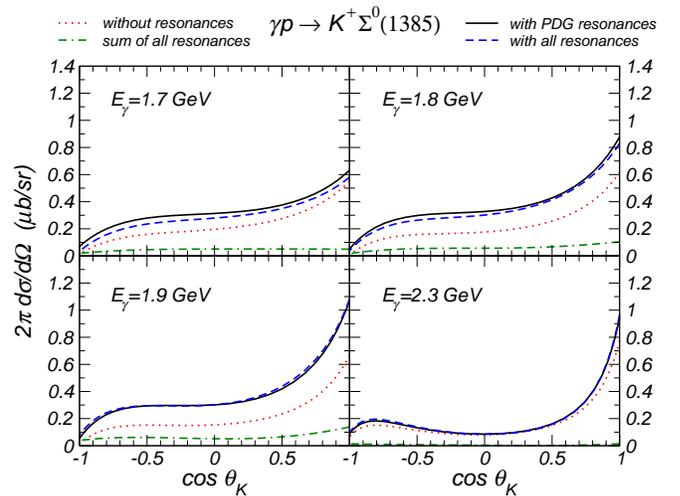}
\caption{(Color online) Differential cross sections for $\gamma p
\to K^+ \Sigma^0(1385)$ at $E_\gamma = 1.7$, $1.8$, $1.9$, and $2.3$
GeV with the inclusion of resonances.
The dash-dash-dotted line is the sum of the resonance terms.
See the text for the details.}
\label{fig:diff-res}
\end{figure}

Similar conclusions on the role of resonances in the reaction
$\gamma p \to K^+ \Sigma^0(1385)$ can be drawn from its differential
cross sections shown in Fig.~\ref{fig:diff-res}. The solid and
dashed lines, which are obtained with the PDG resonances and with
all resonances, respectively, are close to each other, but they can
be distinguished from the dotted lines that are obtained without the
resonant contribution, provided the data are accurate enough, in
particular, in the region of $E_\gamma = 1.8 \sim 1.9$ GeV. At
higher energies, the models with and without resonances give nearly
the same result.  The contributions from the sum of all the
resonances considered in the present work are given by the
dot-dashed lines.

\begin{figure}[tb]
\centering
\includegraphics[width=1.0\hsize,angle=0]{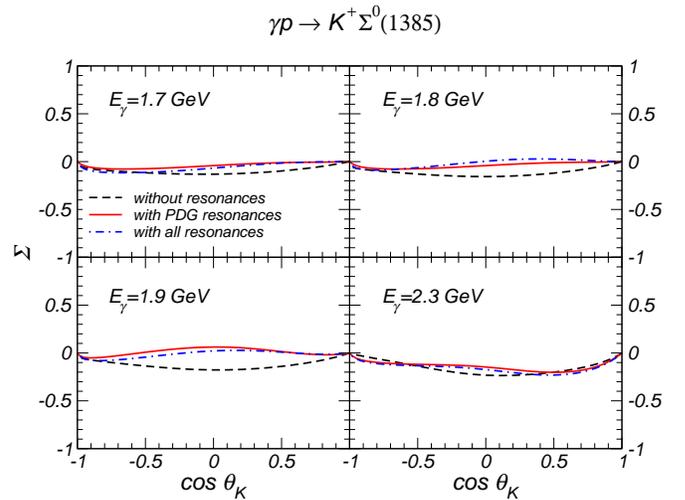}
\caption{(Color online) Single photon asymmetry for the
reaction $\gamma p \to K^+ \Sigma^0(1385)$ at $E_\gamma = 1.7$,
$1.8$, $1.9$, and $2.3$ GeV.
Dashed lines are without the resonance contributions.
Solid lines are obtained with the PDG resonances, whereas the dot-dashed
lines are with all resonances considered in the present work.}
\label{fig:pols-all}
\end{figure}

We next consider the photon single asymmetry in the reaction $\gamma
p \to K^+ \Sigma^0(1385)$, which is defined as
\begin{equation}
\Sigma = \frac{d\sigma_x/d\Omega - d\sigma_y/d\Omega}{d\sigma_x/d\Omega
- d\sigma_y/d\Omega},
\end{equation}
where $d\sigma_x/d\Omega$ and $d\sigma_y/d\Omega$ are the
differential cross sections with linearly polarized photons in the
$x$-direction and in the $y$-direction, respectively.
Here, the $x$-direction and the beam momentum direction (i.e., the
$z$-direction) define the reaction plane, and the $y$-direction is
transverse to the reaction plane.
The results are shown in Fig.~\ref{fig:pols-all}, and it is seen that
the role of the resonances can be verified by measuring the photon
asymmetry.%
\footnote{It should, however, be noted that the final-state
interactions may give nontrivial contributions to spin asymmetries.}
The similarities observed between the solid line (obtained with PDG
resonances) and the dot-dashed line (obtained with all resonances)
make it, however, difficult to identify the role of the missing
resonances.

\section{Summary and Discussion} \label{sec:conclusion}

In this paper, we have studied the reaction mechanisms for
$K\Sigma(1385)$ production in photon-proton collisions.
We find that the peak observed in the preliminary total cross section
data of the CLAS Collaboration requires the inclusion of the resonance
contribution in the production mechanism.
We have accounted for the role of the resonances based on the
effective Lagrangian approach. In the present work, we have
considered eight nucleon and $\Delta$ resonances.
Five of them are listed in PDG (Table~\ref{tab:res-A}) and three of them
are missing resonances predicted by the quark model (Table~\ref{tab:res-B}).
However, the properties of these resonances are poorly known or
unknown even for the PDG resonances, and we have thus
relied on the predictions of hadronic models for the
resonance parameters. In particular, we have related the amplitudes
of $R \to N\gamma$ and $R \to K\Sigma(1385)$ decays with the
coupling constants of our effective Lagrangian, and the predictions
of a quark model made in Refs.~\cite{Caps92,CR98b} for the decay
amplitudes are then used to determine these coupling constants.

The results obtained in this work show that the most important
contribution comes from $\Delta(2000) F_{35}$, and the contributions
of $\Delta(1940) D_{33}$ and $N(2080) D_{13}$ are also important.
Among the missing resonances, the $N\frac32^-(2095)$ contribution is
comparable to those of $\Delta(1940) D_{33}$ and $N(2080) D_{13}$.
Although this resonance has the largest partial decay width into the
$K\Sigma(1385)$ channel, its small photon helicity amplitudes into
$N\gamma$ reduces its contribution to this reaction.
Furthermore, the contributions from the missing resonances are found
to have destructive interference with other missing resonances, and
this makes the sum of the missing resonance terms rather small.
This is also verified by our results on the photon single asymmetry
in this reaction.
Our predictions for the cross section and photon single asymmetry show
a significant difference between the models with and without the
resonances, and this can be verified by experiments at the currently
available electron/photon facilities.

It should be stressed that further works to unravel the properties
of the resonances is strongly required. For example, some other
missing resonances such as $N\frac12^-(2070)$ and $N\frac52^-(2260)$
are predicted in Ref.~\cite{CR98b} to have large couplings to the
$K\Sigma(1385)$ channel. However, we could not study their role in
the reaction $\gamma p \to K^+ \Sigma^0(1385)$ as their photon
helicity amplitudes are unknown in the same quark model.
Furthermore, the properties of resonance decays into $K\Sigma(1385)$
and $N\gamma$ should also be investigated by other models of hadron
structure. This would help us to improve our understanding of the
resonances and search for the missing ones. Finally, to
identify the role of resonances of different isospin, it is
desirable to study
$K\Sigma(1385)$ photoproduction in other isospin channels.

\acknowledgments

We are grateful to L. Guo for useful discussions on the preliminary
CLAS data. Two of us (Y.O. and C.M.K.) are grateful to the
organizers of a focus program, ``Hadron Physics at RHIC'', of the
Asia Pacific Center for Theoretical Physics where this work was
completed. The work of Y.O. and C.M.K. was supported by the US
National Science Foundation under Grant No.\ PHY-0457265 and the
Welch Foundation under Grant No.\ A-1358 while that of K.N. was
supported by Forschungszentrum-J{\"u}lich, under FFE grant No.\
41445282.

\appendix

\section{Propagators and isospin structure}
\label{app:prop}

The propagator of a spin-$3/2$ Rarita-Schwinger field of momentum
$p$ and mass $M$ reads as
\begin{equation}
\Delta_{\alpha\beta}(p,M) = \frac{p\!\!\!/ - M}{p^2 - M^2}
S_{\alpha\beta}(p,M),
\end{equation}
where
\begin{eqnarray}
S_{\alpha\beta}(p,M) &=& -\bar{g}_{\alpha\beta}^{} + \frac13
\bar{\gamma}_\alpha^{} \bar{\gamma}_\beta^{}.
\end{eqnarray}
With
\begin{eqnarray}
\bar{g}_{\mu\nu}^{} &=& g_{\mu\nu}^{} - \frac{p_\mu^{}
p_\nu^{}}{M^2},
\nonumber \\
\bar{\gamma}_\mu^{} &=& \gamma_\mu^{} - \frac{p_\mu^{}}{M^2}
p\!\!\!/,
\end{eqnarray}
this leads to
\begin{eqnarray}
S_{\alpha\beta}(p,M) &=& -g_{\alpha\beta}^{} + \frac13
\gamma_\alpha^{} \gamma_\beta^{} + \frac{1}{3M} \left(
\gamma_\alpha^{} p_\beta^{} - p_\alpha^{} \gamma_\beta^{} \right)
\nonumber \\ && \mbox{}
+ \frac{2}{3M^2}p_\alpha^{} p_\beta^{}.
\end{eqnarray}

The propagator of a spin-$5/2$ baryon of momentum $p$ and mass $M$
is written as~\cite{Rush66,BF57,Chang67a,Scholten}
\begin{equation}
\Delta_{\alpha\beta;\mu\nu}(p,M) = \frac{p\!\!\!/ - M}{p^2 - M^2}
S_{\alpha\beta;\mu\nu}(p,M),
\end{equation}
where
\begin{eqnarray}
S_{\alpha\beta;\mu\nu}(p,M) &=& \frac12 \left(
\bar{g}_{\alpha\mu}^{} \bar{g}_{\beta\nu}^{} +
\bar{g}_{\alpha\nu}^{} \bar{g}_{\beta\mu}^{} \right) - \frac15
\bar{g}_{\alpha\beta}^{} \bar{g}_{\mu\nu}^{} \nonumber \\ && \mbox{}
\hspace{-2cm} - \frac{1}{10} \left( \bar{\gamma}_\alpha^{}
\bar{\gamma}_\mu^{} \bar{g}_{\beta\nu}^{} + \bar{\gamma}_\alpha^{}
\bar{\gamma}_\nu^{} \bar{g}_{\beta\mu}^{} + \bar{\gamma}_\beta^{}
\bar{\gamma}_\mu^{} \bar{g}_{\alpha\nu}^{} + \bar{\gamma}_\beta^{}
\bar{\gamma}_\nu^{} \bar{g}_{\alpha\mu}^{} \right).
\nonumber \\
\end{eqnarray}
For resonances with finite width $\Gamma$, the mass $M$ in the
propagator is replaced by $M - i \Gamma/2$.

Since we are considering nucleon and $\Delta$ resonances, the
resonance field $R$ has either isospin-$1/2$ or isospin-$3/2$. By
omitting the space-time indices, the isospin structure of
$RK\Sigma^*$ interaction reads as
\begin{equation}
\overline{R} \bm{\Sigma}^* \cdot \bm{\tau} K,
\end{equation}
for isospin-$1/2$ resonance $R$. If the resonance $R$ has
isospin-$3/2$, the effective Lagrangian has the isospin structure as
\begin{equation}
\overline{R} \, \bm{T}_{3/2,1/2} \cdot \bm{\Sigma}^* K,
\end{equation}
where
\begin{widetext}
\begin{eqnarray}
T^{(+1)}_{3/2,1/2} = \left( \begin{array}{cc} \sqrt3 & 0 \\ 0 & 1 \\
0 & 0 \\ 0 & 0 \end{array} \right), \quad T^{(0)}_{3/2,1/2} = \left(
\begin{array}{cc} 0 & 0 \\ \sqrt2 & \sqrt2 \\ 0 & 0 \\ 0 & 0
\end{array} \right),
\quad T^{(-1)}_{3/2,1/2} = \left( \begin{array}{cc} 0 & 0 \\ 0 & 0 \\
1 & 0 \\ 0 & \sqrt3 \end{array} \right).
\end{eqnarray}
In the interaction Lagrangians presented in Appendix~\ref{app:amp},
the isospin structure given above is always understood.

\section{Coupling Constants and Decay amplitudes}
\label{app:amp}

The effective Lagrangians for photoexcitation of a resonance from a
nucleon can be written as
\begin{eqnarray}
\mathcal{L}_{RN\gamma}(\textstyle\frac12^\pm) &=& \frac{ef_1}{2M_N}
\bar{N} \Gamma^{(\mp)} \sigma_{\mu\nu} \partial^\nu A^\mu R +
\mbox{H.c.},
\nonumber \\
\mathcal{L}_{RN\gamma}(\textstyle\frac32^\pm) &=& -
\frac{ief_1}{2M_N} \overline{N} \Gamma^{(\pm)}_\nu F^{\mu\nu} R_\mu
- \frac{ef_2}{(2M_N)^2} \partial_\nu \bar{N} \Gamma^{(\pm)}
F^{\mu\nu} R_\mu + \mbox{H.c.},
\nonumber \\
\mathcal{L}_{RN\gamma}(\textstyle\frac52^\pm) &=&
\frac{ef_1}{(2M_N)^2} \bar{N} \Gamma_\nu^{(\mp)} \partial^\alpha
F^{\mu\nu} R_{\mu\alpha} -\frac{ief_2}{(2M_N)^3} \partial_\nu
\bar{N} \Gamma^{(\mp)} \partial^\alpha F^{\mu\nu} R_{\mu\alpha} +
\mbox{H.c.},
\end{eqnarray}
for $j^\pi = \frac12^\pm$, $\frac32^\pm$, and $\frac52^\pm$
resonances. In the above, $A^\mu$ is the photon field with
$F^{\mu\nu} = \partial^\mu A^\nu - \partial^\nu A^\mu$; $R$,
$R_\mu$, and $R_{\mu\nu}$ are the spin-$1/2$, spin-$3/2$, and
spin-$5/2$ resonance fields, respectively; and $\Gamma^{(\pm)}_\mu$
and $\Gamma^{(\pm)}$ are defined in Eq.~(\ref{gamma_pm}).

The coupling constants $f_1$ and $f_2$ in $\mathcal{L}_{RN\gamma}$
are related to the photon helicity amplitudes of the resonance $R$,
which are defined as
\begin{equation}
\Gamma(R \to N\gamma) = \frac{k_\gamma^2}{\pi}
\frac{2M_N}{(2j+1)M_R} \left[ |A_{1/2}|^2 + |A_{3/2}|^2 \right],
\end{equation}
where $k_\gamma = (M_R^2 - M_N^2)/(2M_R)$ and $M_R$ is the resonance
mass. With our effective Lagrangians, the helicity amplitudes are
expressed as~\cite{SLMP04}
\begin{eqnarray}
A_{1/2}(\textstyle\frac12^\pm) &=& \mp \frac{ef_1}{2M_N}
\sqrt{\frac{k_\gamma M_R}{M_N}},
\nonumber \\
A_{1/2}(\textstyle\frac32^\pm) &=& \mp \frac{e\sqrt6}{12}
\sqrt{\frac{k_\gamma}{M_N M_R}} \left[ f_1 + \frac{f_2}{4M_N^2} M_R
(M_R \mp M_N) \right],
\nonumber \\
A_{3/2}(\textstyle\frac32^\pm) &=& \mp \frac{e\sqrt2}{4M_N}
\sqrt{\frac{k_\gamma M_R}{M_N}} \left[ f_1 \mp \frac{f_2}{4M_N}(M_R \mp
M_N) \right],
\nonumber \\
A_{1/2}(\textstyle\frac52^\pm) &=& \pm \frac{e}{4\sqrt{10}}
\frac{k_\gamma}{M_N} \sqrt{\frac{k_\gamma}{M_N M_R}} \left[ f_1 +
\frac{f_2}{4M_N^2} M_R (M_R \pm M_N) \right],
\nonumber\\
A_{3/2}(\textstyle\frac52^\pm) &=& \pm \frac{e}{4\sqrt5}
\frac{k_\gamma}{M_N^2} \sqrt{\frac{k_\gamma M_R}{M_N}} \left[ f_1
\pm \frac{f_2}{4M_N} (M_R \pm M_N) \right],
\end{eqnarray}
where the spin-parity of the resonance is given in parentheses.

For the interaction Lagrangians describing the decay of a
baryon resonance $R$ of spin-parity $j^\pi$ to $K\Sigma(1385)$ of
the spin-parity combination $0^- + \frac32^+$,
consideration of angular momentum and parity conservation leads to only
one term for $j^\pi = \frac12^\pm$ resonances and two terms for
resonances with $j \ge \frac32$.
The general form of these interaction Lagrangians can be written as
\begin{eqnarray}
\mathcal{L}_{RK\Sigma^*}(\textstyle\frac12^\pm) &=& \frac{h_1}{M_K}
\partial_\mu K \bar{\Sigma}^{*\mu} \Gamma^{(\mp)} R + \mbox{H.c.},
\nonumber \\
\mathcal{L}_{RK\Sigma^*}(\textstyle\frac32^\pm) &=& \frac{h_1}{M_K}
\partial^\alpha K \bar{\Sigma}^{*\mu} \Gamma_\alpha^{(\pm)} R_\mu
+ \frac{ih_2}{M_K^2} \partial^\mu \partial^\alpha K
\bar{\Sigma}^*_\alpha \Gamma^{(\pm)} R_\mu + \mbox{H.c.},
\nonumber \\
\mathcal{L}_{RK\Sigma^*}(\textstyle\frac52^\pm) &=& \frac{i
h_1}{M_K^2}
\partial^\mu \partial^\beta K \bar{\Sigma}^{*\alpha} \Gamma_\mu^{(\mp)}
R_{\alpha\beta} - \frac{h_2}{M_K^3} \partial^\mu \partial^\alpha
\partial^\beta K \bar{\Sigma}^*_\mu \Gamma^{(\mp)} R_{\alpha\beta}
+ \mbox{H.c.}.
\end{eqnarray}

The decay width of resonance $R$ into $K\Sigma(1385)$ is then
obtained as
\begin{eqnarray}
\Gamma(\textstyle\frac12^\pm \to K\Sigma^*) &=& \frac{h_1^2}{2\pi}
\frac{q^3 M_R}{M_K^2 M_{\Sigma^*}^2} (E_{\Sigma^*} \pm
M_{\Sigma^*}),
\nonumber \\
\Gamma(\textstyle\frac32^\pm \to K\Sigma^*) &=& \frac{1}{24\pi}
\frac{q}{M_R M_{\Sigma^*}^2} (E_{\Sigma^*} \mp M_{\Sigma^*})
\Biggl\{ \frac{h_1^2}{M_K^2} (M_R \pm M_{\Sigma^*})^2 \left( 2
E_{\Sigma^*}^2 \mp 2 E_{\Sigma^*} M_{\Sigma^*} + 5 M_{\Sigma^*}^2
\right) \nonumber \\ && \qquad\qquad\qquad\mbox{} \mp 2 \frac{h_1
h_2}{M_K^3} M_R q^2 (M_R \pm M_{\Sigma^*}) (2E_{\Sigma^*} \mp
M_{\Sigma^*}) + 2 \frac{h_2^2}{M_K^4} M_R^2 q^4 \Biggr\},
\nonumber \\
\Gamma(\textstyle\frac52^\pm \to K\Sigma^*) &=& \frac{1}{60\pi}
\frac{q^3}{M_R M_{\Sigma^*}^2} (E_{\Sigma^*} \pm M_{\Sigma^*})
\Biggl\{ \frac{h_1^2}{M_K^4} (M_R \mp M_{\Sigma^*})^2 \left( 4
E_{\Sigma^*}^2 \pm 4 E_{\Sigma^*} M_{\Sigma^*} + 7
M_{\Sigma^*}^2\right)
\nonumber \\
&& \qquad\qquad\qquad\mbox{} \mp 4 \frac{h_1 h_2}{M_K^5} M_R q^2
(M_R \mp M_{\Sigma^*}) (2E_{\Sigma^*} \pm M_{\Sigma^*}) + 4
\frac{h_2^2}{M_K^6} M_R^2 q^4 \Biggr\}, \label{eq:R-decay}
\end{eqnarray}
depending on its spin-parity. In the above, $q$ is the magnitude of
the three-momenta of final-state particles in the rest frame of the
resonance,
\begin{equation}
q = \frac{1}{2M_R} \sqrt{\left[M_R^2 - (M_{\Sigma^*} + M_K)^2
\right] \left[ M_R^2 - (M_{\Sigma^*} - M_K)^2 \right]},
\end{equation}
and $E_{\Sigma^*} = \sqrt{M_{\Sigma^*}^2 + q^2}$. Above formulas are
valid for the decays of resonances of isospin-$1/2$ as well as
isospin-$3/2$.

For a $j=\frac12$ resonance, the decay width can be used to
determine the magnitude of the coupling constant $h_1$ but not its
phase. For resonances with $j \ge \frac32$, this gives only one
relation for two coupling constants, $h_1$ and $h_2$. Therefore, we
need to know the decay amplitudes to uniquely determine the coupling
constants. The signs of the couplings are then fixed by hadron model
predictions.

The decay amplitude for $R \to K\Sigma(1385)$ can be written as
\begin{equation}
\langle K(\bm{q})\, \Sigma^*(-\bm{q},m_f^{}) | -i\, \mathcal{H}_{\rm
int} | R (\bm{0}, m_j^{}) \rangle = 2\pi M_R \sqrt{\frac{2}{q}}
\sum_{\ell,m_\ell^{}} \langle \ell\, m_\ell^{}\, \textstyle\frac32\,
m_f^{} | j \, m_j^{} \rangle\, Y_{\ell m_\ell^{}}(\hat{\bm q})
G(\ell),
\end{equation}
where $Y_{\ell m_\ell^{}}(\hat{\bm q})$ and $\langle \ell\,
m_\ell^{}\, \textstyle\frac32\, m_f^{} | j \, m_j^{} \rangle$ are
the spherical harmonics and Clebsch-Gordan
coefficient, respectively. This also defines the partial wave decay
amplitude $G(\ell)$. The relative orbital angular momentum $\ell$ of
the final state is constrained by the spin-parity of the resonance.
The decay width is then given by
\begin{equation}
\Gamma(R \to K\Sigma^*) = \sum_{\ell} |G(\ell)|^2, \label{eq:Gell}
\end{equation}
where the values of $G(\ell)$ can be obtained from the prediction of
hadronic models, for example, in Ref.~\cite{CR98b}.

For the decay of a $j^\pi = \frac12^\pm$ resonance, angular momentum
conservation restricts the relative orbital angular momentum to
$\ell = 1, 2$. For the decay of a positive parity resonance, the
final-state particles are therefore in the relative $p$ wave, while
they are in the relative $d$-wave in the decay of a negative parity
resonance. In this case, we have
\begin{equation}
G(1) = - \frac{1}{\sqrt{2\pi}} \frac{q}{M_{\Sigma^*}} \sqrt{qM_R}
\sqrt{E_{\Sigma^*} + M_{\Sigma^*}} \frac{h_1}{M_K},
\end{equation}
for a $j^\pi = \frac12^+$ resonance, and
\begin{equation}
G(2) = - \frac{1}{\sqrt{2\pi}} \frac{q}{M_{\Sigma^*}} \sqrt{qM_R}
\sqrt{E_{\Sigma^*} - M_{\Sigma^*}} \frac{h_1}{M_K},
\end{equation}
for a $j^\pi = \frac12^-$ resonance.

For a resonance of $j=\frac32$, the final $K\Sigma(1385)$ state is
in the relative $p$ and $f$ waves in the decay of a positive parity
resonance and are in the relative $s$ and $d$ waves in the decay of
a negative parity resonance. The decay amplitudes can be written in
terms of the coupling constants as
\begin{eqnarray}
G(1) &=& G_{11}^{(3/2)} \frac{h_1}{M_K} + G_{12}^{(3/2)}
\frac{h_2}{M_K^2},
\nonumber \\
G(3) &=& G_{31}^{(3/2)} \frac{h_1}{M_K} + G_{32}^{(3/2)}
\frac{h_2}{M_K^2},
\end{eqnarray}
for a positive parity resonance, where
\begin{eqnarray}
G_{11}^{(3/2)} &=& \frac{\sqrt{30}}{60\sqrt{\pi}}
\frac{1}{M_{\Sigma^*}} \sqrt{\frac{q}{M_R}} \sqrt{E_{\Sigma^*} -
M_{\Sigma^*}} (M_R + M_{\Sigma^*}) (E_{\Sigma^*} + 4 M_{\Sigma^*}),
\nonumber \\
G_{12}^{(3/2)} &=& -\frac{\sqrt{30}}{60\sqrt{\pi}} \frac{q^2 \sqrt{q
M_R}}{M_{\Sigma^*}} \sqrt{E_{\Sigma^*} - M_{\Sigma^*}}
\nonumber \\
G_{31}^{(3/2)} &=& -\frac{\sqrt{30}}{20\sqrt{\pi}}
\frac{1}{M_{\Sigma^*}} \sqrt{\frac{q}{M_R}} \sqrt{E_{\Sigma^*} -
M_{\Sigma^*}} (M_R + M_{\Sigma^*}) (E_{\Sigma^*} - M_{\Sigma^*}),
\nonumber \\
G_{32}^{(3/2)} &=& \frac{\sqrt{30}}{20\sqrt{\pi}} \frac{q^2 \sqrt{q
M_R}}{M_{\Sigma^*}} \sqrt{E_{\Sigma^*} - M_{\Sigma^*}},
\end{eqnarray}
and
\begin{eqnarray}
G(0) &=& G_{01}^{(3/2)} \frac{h_1}{M_K} + G_{02}^{(3/2)}
\frac{h_2}{M_K^2},
\nonumber \\
G(2) &=& G_{21}^{(3/2)} \frac{h_1}{M_K} + G_{22}^{(3/2)}
\frac{h_2}{M_K^2},
\end{eqnarray}
for a negative parity resonance, where
\begin{eqnarray}
G_{01}^{(3/2)} &=& \frac{\sqrt{6}}{12\sqrt{\pi}}
\frac{1}{M_{\Sigma^*}} \sqrt{\frac{q}{M_R}} \sqrt{E_{\Sigma^*} +
M_{\Sigma^*}} (M_R - M_{\Sigma^*}) (E_{\Sigma^*} + 2 M_{\Sigma^*}),
\nonumber \\
G_{02}^{(3/2)} &=& \frac{\sqrt{6}}{12\sqrt{\pi}} \frac{q^2 \sqrt{q
M_R}}{M_{\Sigma^*}} \sqrt{E_{\Sigma^*} + M_{\Sigma^*}}
\nonumber \\
G_{21}^{(3/2)} &=& -\frac{\sqrt{6}}{12\sqrt{\pi}}
\frac{1}{M_{\Sigma^*}} \sqrt{\frac{q}{M_R}} \sqrt{E_{\Sigma^*} +
M_{\Sigma^*}} (M_R - M_{\Sigma^*}) (E_{\Sigma^*} - M_{\Sigma^*}),
\nonumber \\
G_{22}^{(3/2)} &=& -\frac{\sqrt{6}}{12\sqrt{\pi}} \frac{q^2 \sqrt{q
M_R}}{M_{\Sigma^*}} \sqrt{E_{\Sigma^*} + M_{\Sigma^*}}.
\end{eqnarray}

In the decay of a spin-$5/2$ resonance into $K\Sigma(1385)$, the
final state is in the relative $p$ and $f$ waves for a positive
parity resonance and is in the relative $d$ and $g$ waves for a
negative parity resonance. The decay amplitudes are then written as
\begin{eqnarray}
G(1) &=& G_{11}^{(5/2)} \frac{h_1}{M_K^2} + G_{12}^{(5/2)}
\frac{h_2}{M_K^3},
\nonumber \\
G(3) &=& G_{31}^{(5/2)} \frac{h_1}{M_K^2} + G_{32}^{(5/2)}
\frac{h_2}{M_K^3},
\end{eqnarray}
for a positive parity resonance, where
\begin{eqnarray}
G_{11}^{(5/2)} &=& - \frac{1}{10\sqrt{\pi}} \frac{q}{M_{\Sigma^*}}
\sqrt{\frac{q}{M_R}} \sqrt{E_{\Sigma^*} + M_{\Sigma^*}} (M_R -
M_{\Sigma^*})(2 E_{\Sigma^*} + 3 M_{\Sigma^*}),
\nonumber \\
G_{12}^{(5/2)} &=& \frac{1}{5\sqrt{\pi}} \frac{q^3
\sqrt{qM_R}}{M_{\Sigma^*}} \sqrt{E_{\Sigma^*} + M_{\Sigma^*}},
\nonumber \\
G_{31}^{(5/2)} &=& - \frac{\sqrt6}{15\sqrt{\pi}}
\frac{q}{M_{\Sigma^*}} \sqrt{\frac{q}{M_R}} \sqrt{E_{\Sigma^*} +
M_{\Sigma^*}} (M_R - M_{\Sigma^*})(E_{\Sigma^*} - M_{\Sigma^*}),
\nonumber \\
G_{32}^{(5/2)} &=& -\frac{\sqrt6}{15\sqrt{\pi}} \frac{q^3
\sqrt{qM_R}}{M_{\Sigma^*}} \sqrt{E_{\Sigma^*} + M_{\Sigma^*}},
\end{eqnarray}
and
\begin{eqnarray}
G(2) &=& G_{21}^{(5/2)} \frac{h_1}{M_K^2} + G_{22}^{(5/2)}
\frac{h_2}{M_K^3},
\nonumber \\
G(4) &=& G_{41}^{(5/2)} \frac{h_1}{M_K^2} + G_{42}^{(5/2)}
\frac{h_2}{M_K^3},
\end{eqnarray}
for a negative parity resonance, where
\begin{eqnarray}
G_{21}^{(5/2)} &=& - \frac{\sqrt{105}}{210\sqrt{\pi}}
\frac{q}{M_{\Sigma^*}} \sqrt{\frac{q}{M_R}} \sqrt{E_{\Sigma^*} -
M_{\Sigma^*}} (M_R + M_{\Sigma^*})(2 E_{\Sigma^*} + 5 M_{\Sigma^*}),
\nonumber \\
G_{22}^{(5/2)} &=& -\frac{\sqrt{105}}{105\sqrt{\pi}} \frac{q^3
\sqrt{qM_R}}{M_{\Sigma^*}} \sqrt{E_{\Sigma^*} - M_{\Sigma^*}},
\nonumber \\
G_{41}^{(5/2)} &=& \frac{\sqrt{70}}{35\sqrt{\pi}}
\frac{q}{M_{\Sigma^*}} \sqrt{\frac{q}{M_R}} \sqrt{E_{\Sigma^*} -
M_{\Sigma^*}} (M_R + M_{\Sigma^*})(E_{\Sigma^*} - M_{\Sigma^*}),
\nonumber \\
G_{42}^{(5/2)} &=& \frac{\sqrt{70}}{35\sqrt{\pi}} \frac{q^3
\sqrt{qM_R}}{M_{\Sigma^*}} \sqrt{E_{\Sigma^*} - M_{\Sigma^*}}.
\end{eqnarray}

\end{widetext}

It can be verified that the decay widths obtained from
Eq.~(\ref{eq:Gell}) with the above relations reproduce the results
given in Eq.~(\ref{eq:R-decay}). Above effective Lagrangians are
constructed for resonances of spin up to $5/2$, but they can be
straightforwardly generalized to resonances with arbitrary
spin-parity.

\end{document}